\documentstyle[11pt]{article}

\title{\bf Estimates of the subthreshold photoproduction of charm}
\author{M.A.Braun \thanks{Permanent address: 
Dep. High-Energy physics,
S.Petersburg University, 198504 S.Petersburg, Russia}
\hspace{2 mm} and B.Vlahovic\\
NCCU,Durham, NC, USA}
\date{}
\def\beq{\begin{equation}}
\def\eeq{\end{equation}}

\begin{document}
\maketitle
\medskip
\vspace{1 cm}

{\bf Abstract}
\vspace{1 cm}

Charm photoproduction rates off nuclei below the nucleon threshold are estimated
using the phenomenologically known structure functions both for $x<1$ and $x>1$.
The rates rapidly fall below the threshold from values  $\sim$1 nb for Pb at the
threshold (7.8 GeV) to $\sim$1pb at 6 GeV.

\section{Introduction}
In view of the envisaged upgrade of the CEBAF facility up to 12 GeV
it becomes important to have relatively secure predictions about the
production rates of charm on nuclear targets below and immediately 
above the threshold for the nucleon target. This note aims at such
predictions. From the start it has to be stressed that the dynamical
picture of charm production at energies close to the threshold is
much more complicated than at high energies (see e.g. [1]).
Correspondingly the production rates cannot be described by the
standard collinear factorization expression but involve 
gluon distributions both in $x$ and transverse momentum, which are
unknown at small transverse momenta (in the confinement region).
To obtain our predictions we shall recur to some crude assumptions
about these distributions both at $x<1$ and $x>1$. In the latter
case we shall exploit the (poorly) known behaviour of the nuclear 
structure functions, which fall exponentially with $x$, the slope 
being in the range 15$\div$16. With all these simplifications, we
hope to be able to predict the rates up to factor 2$\div$3.

\section{Kinematics}
Consider the exclusive process shown in Fig. 1
\beq
\gamma+A\to C{\bar C}+A^*
\eeq
where $A$ is the target nucleus of mass $m_A$ and
$A^*$ is the recoil nuclear system of mass $m_A^*$. The notation for all 
the momenta
involved can be read off the Fig. 1. We denote the total mass of the
$C\bar C$ 
system as $M$.
Obviously $M\geq 2m_c$ where $m_c$ is the mass of the C-quark, which for our
modest purpose we take as 1.5 GeV. The inclusive cross-section
for charm photoproduction is obtained after summing over all states of the
recoil nuclear system.

We choose a reference system in which the target nucleus is at rest and 
the incoming photon is
moving along the $z$-axis in the opposite direction, so that $q_+=q_{\perp}=0$ 
The (part of) the photoproduction cross-section corresponding to (1) is 
then obtained via the
imaginary part of the diagram in Fig. 1 as
\beq
\sigma_A=A\int\frac{d^4k}{(2\pi)^3}\delta((Ap-k)^2-{m_a^*}^2)x
\left(\frac{\Gamma(k^2)}{k^2}\right)^2\sigma_g(M^2,k^2)
\eeq
Here $\Gamma(k^2)$ is the vertex for gluon emission from the target; 
$\sigma_g(M^2,k^2)$
is the photoproduction cross-section off the virtual gluon of momentum $k$.
We have also introduced the scaling variable for the gluon as $x=k_+/p_+$. 
Due to
$q_+=0$ this is also the scaling variable for the observed charm. Note that 
this
definition, which is standard at large eneries and produced masses, is not 
at all standard
at moderate scales. In particular this $x$ does not go to unity at the 
threshold
for the nucleon target. Rather the limits for its variation converge 
to a common
value 0.76. For the nuclear targets with $A>>1$ its minimal value at the 
nucleon threshold is 
well below unity ($\sim$ 0.64). One should have this in mind when
associating this $x$
with the gluon distribution: it follows that for a nuclear target, for 
energies going 
noticeably below the nucleon threshold, the contributing values of $x$ 
still remain
below $x=1$. This effect is essential for the  charm cross-sections
to remain not too small immediately  below the threshold.

We use the $\delta$-function to integrate over $k_-$ to obtain
imaginary part of the diagram in Fig. 1 as
\beq
\sigma_A=A\int\frac{dxd^2k_{\perp}}{2(A-x)(2\pi)^3}
\left(\frac{\Gamma(k^2)}{k^2}\right)^2\sigma_g(M^2,k^2)
\eeq
In these variables we find
\beq
M^2=xs_1+k^2,\ \ s_1=2pq
\eeq
and
\beq
k^2=xAm^2-\frac{x}{A-x}{m_A^*}^2-\frac{A}{A-x}k^2_{\perp}
\eeq
where we have put $p^2=m^2$, the nucleon mass squared, neglecting the
binding.

The limits of integration in (3) are determined by the condition $M^2\geq
4m_c^2$,
 which leads to
\beq
x(s_1+Am^2)-\frac{x}{A-x}{m_A^*}^2-\frac{A}{A-x}k^2_{\perp}-4m_c^2\geq 0
\eeq
Since $k_{\perp}^2\geq 0$ one gets
\beq
x(s_1+Am^2)-\frac{x}{A-x}{m_A^*}^2-4m_c^2\geq 0
\eeq
from which one finds the limits of integration in $x$:
\beq
x_1\leq x\leq x_2
\eeq
where
\beq
x_{1,2}=\frac{1}{2s}(As-{m_A^*}^2+4m_c^2\pm\sqrt{[As-(m^*+2m_c)^2]
[As-(m^*-2m_c)^2]}
\eeq
where $s=s_1+Am^2$.
The limits of integration in $k_{\perp}$ at a given $x$ are determined by
(6).

Using (5) we may pass from the integration variable $k_{\perp}^2$ to
$|k^2|$:
\beq
\sigma_A=\pi\int_{x_1}^{x_2}xdx\int_{|k^2|_{min}}^{xs_1-4m_c^2}\frac
{d|k^2|}{2(2\pi)^3}
\left(\frac{\Gamma(k^2)}{k^2}\right)^2\sigma_g(xs_1-|k^2|,k^2)
\eeq
where
\beq
|k^2|_{min}=\frac{A}{A-x}x^2m^2
\eeq

The threshold energy corresponds to $x_1=x_2$ or $As=m_A^*+2m_c$ with the 
minimal
possible value of the recoiling system, which is just $m_A\simeq Am$. In 
terms of the
photon energy $E$ we have $s_1=2mE$ and the threshold energy is found to
be
\beq
E^{th}_A=2m_c\left(1+\frac{1}{A}\frac{m_c}{m}\right)
\eeq
It steadily falls with $A$ from the nucleon target threshold
\beq
E^{th}_1=2m_c\left(1+\frac{m_c}{m}\right)\simeq 7.8\ {\rm GeV} 
\eeq
down to the value $2m_c$ for infinitely heavy nucleus.

\section{High-energy limit}
To interprete the quantities enetering Eq. (10) it is instructive to study
its high-energy limit, which corresponds to taking  $s_1 >>m_c^2$ and both 
quantities much greater than
the nucleon mass. Assuming that the effective values of the gluon
virtuality are 
limited
(and small) one then gets for the nucleon target ($A=1$)
\beq
\sigma_1= 
\pi\int_{x_1}^{x_2}xdx\int_{0}^{xs_1}\frac {d|k^2|}{2(2\pi)^3}
\left(\frac{\Gamma(k^2)}{k^2}\right)^2\sigma_g(xs_1)
\eeq
Here we also neglect the off-mass-shellness of the cross-section off the 
gluon, considering
$|k^2|<<4m_c^2$.
The obtained formula is precisely the standard collinear factorization formula
with the identification
\beq
xg(x,M^2)=\pi\int_{0}^{M^2}\frac {d|k^2|}{2(2\pi)^3}
\left(\frac{\Gamma(k^2)}{k^2}\right)^2\equiv\int_{0}^{M^2}d|k^2|x\rho(x,|k^2|)
\eeq
Of course it is understood that this is only a part of the gluon distribution 
in $x$
coming from a particular recoil state. The total gluon distribution is 
 obtained after
summing over all recoiling states.
The quantity $\rho(x,|k^2|)$ obviously has a meaning of the double distribution 
of gluons in $x$ and $|k^2|$. Eq.(15) allows to relate function
$\Gamma^2/k^4$ entering our formulas of Section 2 to the double distribution 
$\rho(x,|k^2|)$
and rewrite our formula for the cross-section at finite energies as
\beq
\sigma_A=\int_{x_1}^{x_2}xdx\int_{|k^2|_{min}}^{xs_1-4m_c^2} d|k^2|
\rho_A(x,|k^2|)\sigma_g(xs_1-|k^2|,k^2)
\eeq

\section{The gluon distribution $\rho(x,|k^2|)$}
To make our estimates, we assume a simple factorizable form for the double 
density
$\rho(x,|k^2|)$ and choose the $|k^2|$ dependence in accordance to the 
perturbation theory
with an infrared cutoff in the infrared region:
\beq
\rho(x,|k^2|)=\frac{a(x)}{|k^2|+\Lambda^2}
\eeq
Function $a(x)$ can be obtained matching (17) with the observed
$xg(x,M^2)$ at a particular point
$M_0^2$. Since we are interested in the threshold region, we take 
$M_0=2m_c$ to finally obtain
\beq
\rho(x,|k^2|)=\frac{g(x,4m_c^2)}{\ln(4m_c^2/\Lambda^2+1)}
\ \frac{1}{|k^2|+\Lambda^2}
\eeq
Our calculations show that the results are practically independent of the
infrared cutoff for energies abov 6 GeV. However for lower energies they 
begin do
rapidly fall with the growth of $\Lambda$. 

For the proton at $x<1$ the gluon distribution $g(x,4m_c^2)$ can be taken from 
numerous existing fits to the experimental data. In our calculations we
have used GRV95 LO [2]. For the nucleus at $x<1$ we
 use 
the simplest assumption $g_A(x,Q^2)=Ag_1(x,Q^2)$ neglecting the EMC effect 
in the first
approximation. Obviously this not a very satisfactory  approximation at
$x$ quite close to unity.

For the nuclei at  $x>1$ we  the gluon distribution
may be estimated 
using, first, the existing data for the structure function in this region
and, second, the popular hypothesis that at sufficiently low $Q^2=s_0$ the 
gluon 
distribution vanishes and the hadrons become constructed exclusively of quarks.
Then one can, in principle, find the gluon distribution at a given $Q^2$
from the standard DGLAP evolution equation with the quark distributions
determined from the experimental data on the structure functions at $x>1$
evolved back to $Q^2=s_0$. In the present estimates we recurred to a simpler
approach, neglecting the $Q^2$ dependence of the quark distributions at $x>1$
in accordance with the naive parton model and experimental evidence for the
structure functions at $x>1$ which depend on $Q^2$ very weakly.
From [3] one concludes that at $x>1$ 
the structure function
of carbon falls with $x$ as approximately $\exp(-\Delta x)$ with  
$\Delta\simeq 16$,
its value at $x=1.05$ being $\sim 6*10^{-5}$ and very weakly dependent on
$Q^2$.
From these data one can find the quark density at $x>1$ and using the DGLAP 
equation
find the corresponding gluon density at $x>1$:
\beq
xg_A(x,Q^2)=a\frac{6\alpha_s}{5\pi\Delta}A^{1+0.3x}
\frac{1}{x}e^{-\Delta x}\ln{Q^2}{s_0}
\eeq
with $\delta=16.$, $a=100.$ and $s_0$ the scale at which 
the gluon contents of the  hadrons  vanishes. We take $s_0=$(0.4 GeV)$^2$.
The $A$-dependence has been chosen in accordance with the experimental data for
hadron production at $x>1$ [4].

\section{Numerical results}
To simplify  calculations we take in (16) the photon-gluon fusion
cross-section on the gluon mass shell  where it is known to be [5]
\beq
\sigma_g(M^2)=\pi\alpha_{em}\alpha_s e_c^2\frac{1}{M^4}
\int_{t_1}^{t_2} dt\Big[\frac{t}{u}+\frac{u}{t}+\frac{4m_c^2M^2}{tu}
\left(1-\frac{m_c^2M^2}{tu}\right)\Big]
\eeq
where $e_c$ is the quark charge in units $e$, $u=-M^2-t$ and the limits
$t_{1,2}$ are given by
\beq
t_{1,2}=-\frac{M}{2}[M\pm\sqrt{M^2-4m_c^2}]
\eeq
We take the strong coupling constant $\alpha_s=0.3$, infrared cutoff in the
gluon distribution $\Lambda=0.3$ GeV.
 
With these parameters we obtain the cross-sections for charm photoproduction
shown in Fig. 2. To have the idea of the number of nucleons which have to 
interact
together to produce charm at fixed energy below threshold we show the limits of
intergation $x_1$ and $x_2$ in Fig. 3.

As expected the cross-sections rapidly fall for energies below threshold.
Their energy dependence cannot be fit with a simple exponential (in fact
they fall faster than the exponential). As to the aboslute values, for Pb at
we have obtained values around 1 nb at the nucleon threshold (7.8 GeV)  
and around 1 pb at 6 GeV. The $A$ dependence is close to linear.

\section{Discussion}
We have crudely estimated charm photoproduction rates for nuclear
targets below and in the vicinity of the nucleon target threshold.
The estimates require knowledge of the gluon distribution in both $x$ and 
$k_{\perp}^2$ in a wide region of the momenta including the confinement region.

Our estimates were based on a simple factorization assumption and introduction
of an infrared cutoff. As mentioned at energies above 6 GeV the obtained 
cross-sections practically do not change at all when the cutoff is raised from
0.3 GeV to 0.4 GeV. However at 5 GeV they fall by nearly 30 times. Thus at
energies much below the nucleon threshold and closer to the nuclear thresholds
the results become strongly dependent on the cutoff. However at such energies
the cross-sections becomes so small that do not present any interest from the
practical point of view.

The behaviour in the nearest vicinity of the threshold depends crucially
on the choice of the scaling variable $x$ and the form of the nuclear 
gluon distribution 
at $x$ smaller but very close to unity. Our choice of $x$ as the part of 
the "+"
component of the hadron momentum carried by the gluon is well founded from the
theoretical point of view. However with this definition $x$ remains below unity
at the nucleon target threshold. This considerably enhances the cross-sections
for nuclear targets in the vicinity and somewhat below the threshold.  
Our assumption that in this region the nuclear structure function is roughly
proportional to A is quite crude without any doubt. To improve  the 
estimates one has
to specially investigate this region of $x$ for a nuclear target following [1].
  
\section{Acknowledgements}

M.A.B. is thankful to the Faculty of Science of the NCCU
for hospitality. 

\section{References}

1. S.J.Brodsky, E.Chudakov, P.Hoyer and J.M.Laget,
Phys. Lett. {\bf B 498} (2001) 23.

2. M.Glueck, E.Reya and A.Vogt, Z.Phys. {\bf C 67} (1995) 433.

3. BCDMS collab., Z.Phys. {\bf C 63} (1994) 29.

4. Y.D.Bayukov {\it et al.}, Phys. Rev {\bf C 20} (1979) 764;
N.A.Nikiforov {\it et al.}, Phys. Rev. {\bf C 22} (1980) 700.

5. J.Smith and W.L. van Neerven, Nucl. Phys. {\bf B 374} (1992) 36. 

\section{Figure captions}

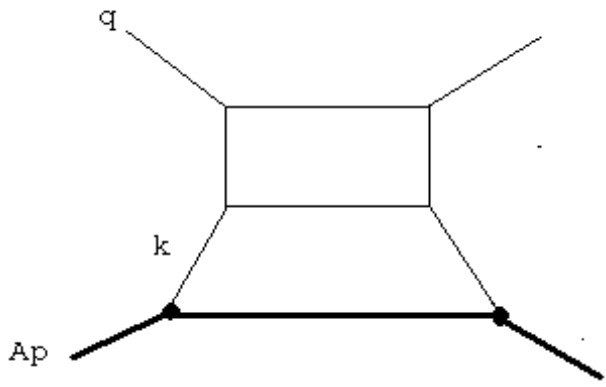
Fig. 1. The forward scattering amplitude corresponding to reaction 
(1).

Fig. 2. The charm photoproduction cross-sections for different photon
energies $E$ and targets. Curves from bottom to top correspond to 
$A$ =1, 12, 64 and 207.

Fig. 3. The limits of $x$-integration for different photon energies and
nuclear targets. Curves from bottom to top correspond to 
$A$ =12, 64 and 207.

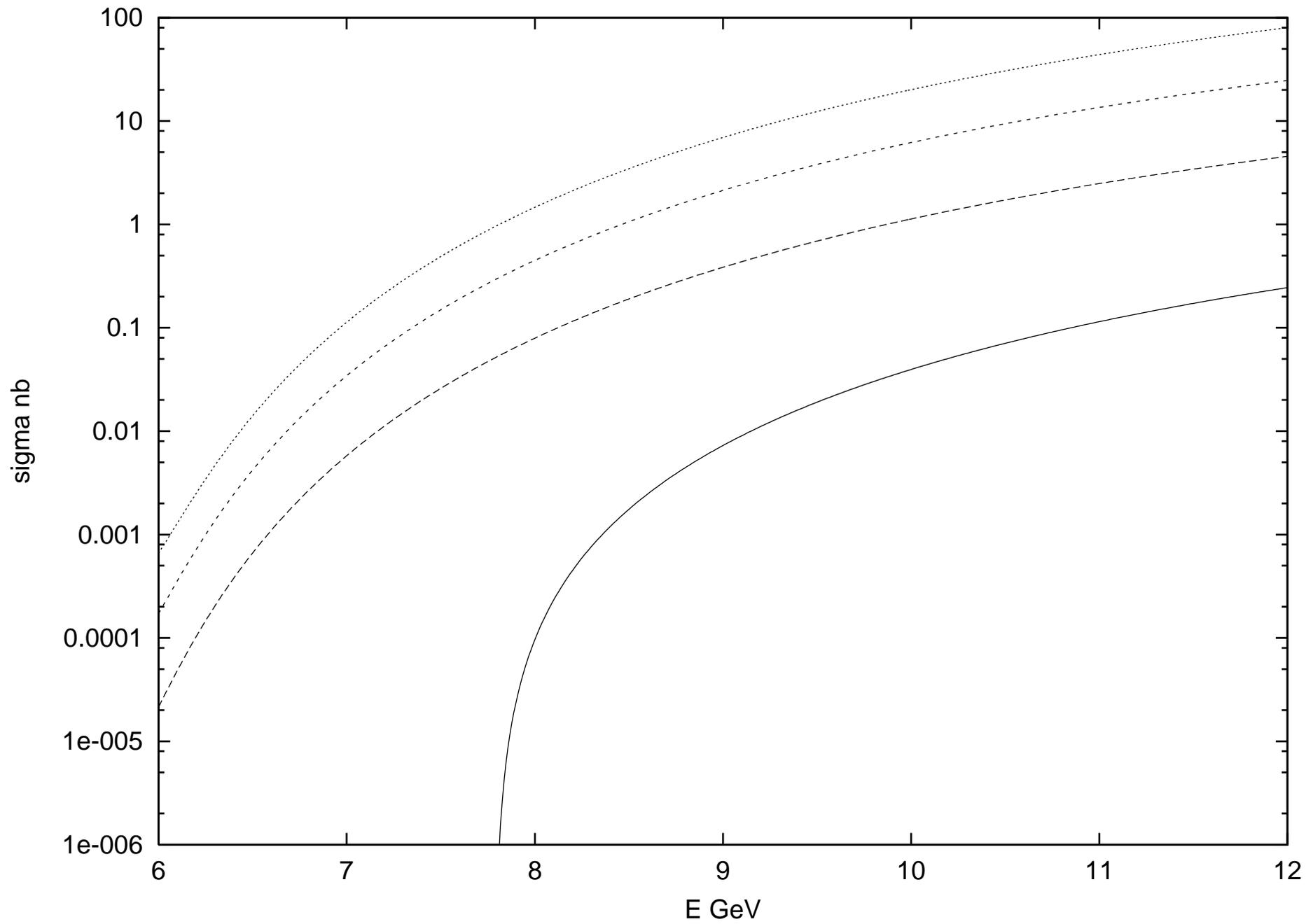
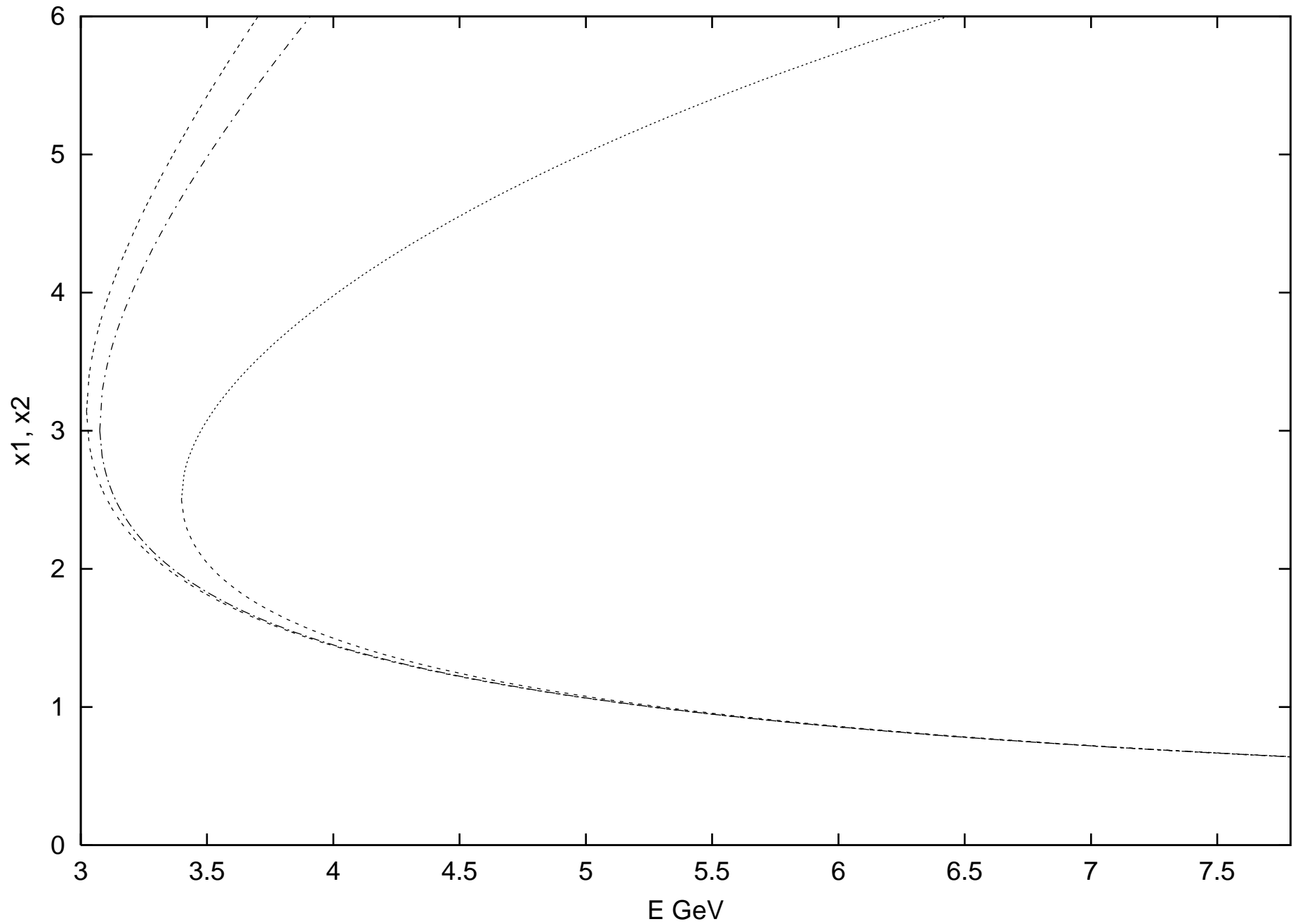
\end{document}